\documentclass[twocolumn,english,showpacs,preprintnumbers,groupedaddress,amsmath,amssymb,aps,prl]{revtex4}
\usepackage[T1]{fontenc}
\usepackage[latin9]{inputenc}
\usepackage{float}
\usepackage{textcomp}
\usepackage{amsmath}
\usepackage{graphicx}
\usepackage{amssymb}
\usepackage{esint}

\makeatletter
\@ifundefined{textcolor}{}
{%
 \definecolor{BLACK}{gray}{0}
 \definecolor{WHITE}{gray}{1}
 \definecolor{RED}{rgb}{1,0,0}
 \definecolor{GREEN}{rgb}{0,1,0}
 \definecolor{BLUE}{rgb}{0,0,1}
 \definecolor{CYAN}{cmyk}{1,0,0,0}
 \definecolor{MAGENTA}{cmyk}{0,1,0,0}
 \definecolor{YELLOW}{cmyk}{0,0,1,0}
 }

\makeatother

\usepackage{babel}

\begin{document}

\title{Comment on ``Optical precursors in the singular and weak dispersion
limits''}

\author{Bruno Macke}

\author{Bernard S\'{e}gard}

\email{bernard.segard@univ-lille-1.fr}

\affiliation{Laboratoire de Physique des Lasers, Atomes et Mol\'{e}cules , CNRS et
Universit\'{e} Lille 1, 59655 Villeneuve d'Ascq, France}

\begin{abstract}
We point out inconsistencies in the recent paper by Oughstun \emph{et
al}. on Sommerfeld and Brillouin precursors {[}J. Opt. Soc. Am. B
\textbf{27}, 1664-1670 (2010){]}. Their study is essentially numerical
and, for the parameters used in their simulations, the difference
between the two limits considered is not as clear-cut as they state.
The steep rise of the Brillouin precursor obtained in the singular
limit and analyzed as a distinguishing feature of this limit simply
results from an unsuitable time scale. In fact, the rise of the precursor
is progressive and is perfectly described by a Airy function. In the
weak dispersion limit, the equivalence relation, established at great
length in Section 3 of the paper, appears as an immediate result in
the retarded-time picture. Last but not least, we show that, contrary
to the authors claim, the precursors are catastrophically affected
by the rise-time of the incident optical field, even when the latter
is considerably faster than the medium relaxation time.

\end{abstract}
\maketitle
In a recent paper \cite{ou10}, Oughstun \emph{et al}. revisit the
classical problem of the propagation of a step modulated pulse in
a Lorentz model medium. They specifically consider the case where
the absorption line is narrow (singular limit) and the one where the
refractive index of the medium keeps very close to unity at every
frequency (weak dispersion limit). The medium is characterized by
its complex refractive index
\begin{equation}
n(\omega)=\left(1-\frac{\omega_{p}^{2}}{\omega^{2}-\omega_{0}^{2}+2i\delta\omega}\right)^{1/2}.\label{eq:1}\end{equation}
 Here $\omega$, $\omega_{p}$, $\omega_{0}$ and $\delta$ respectively
designate the current optical frequency, the plasma frequency, the
resonance frequency and the damping or relaxation rate. The wave propagates
in the $z$-direction. In the following we use the retarded time $t$,
equal to the real time minus $z/c$ where $c$ is the velocity of
light in vacuum (retarded time picture). The medium is then characterized
by the transfer function
\begin{equation}
H(\omega)=\exp\left[i\frac{\omega}{c}z\left(n(\omega)-1\right)\right]\label{eq:2}
\end{equation}
 and the field transmitted $E(z,t)$ at the abscissa $z$ reads as
\begin{equation}
E(z,t)=\frac{1}{2\pi}\int_{-\infty+ia}^{+\infty+ia}H(\omega)\widetilde{E}(0,\omega)\exp(-i\omega t)\mathrm{d}\omega.\label{eq:3}
\end{equation}
 Here $a$ is a positive constant and  $\widetilde{E}(0,\omega)$ is the
Fourier transform of the incident field $E(0,t)$. In \cite{ou10},
the latter is assumed to have the idealized form\begin{equation}
E(0,t)=\Theta(t)\sin(\omega_{c}t)\label{eq:4}\end{equation}
 where $\Theta(t)$ is the unit step function and $\omega_{c}$ is
the frequency of the optical carrier. 

\begin{figure}[h]
\begin{centering}
\includegraphics[width=80mm]{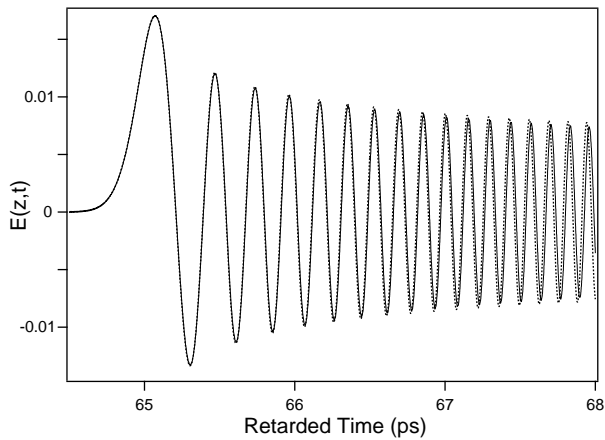} 
\par\end{centering}

\caption{Brillouin precursor obtained for the parameters of the figure 3 of
\cite{ou10}, that is for $\omega_{0}=3.9\times10^{14}\mathrm{rad/s}$,
$\omega_{c}=3.0\times10^{14}\mathrm{rad/s}$, $\omega_{p}=3.05\times10^{14}\mathrm{rad/s}$
, $\delta=3.02\times10^{10}\mathrm{rad/s}$, and $z=7.232\times10^{-2}\mathrm{m}$.
The full line is the exact numerical solution obtained by fast Fourier
transform (FFT) and the dashed line is the approximate analytical
solution given by Eq.\ref{eq:7}. \label{fig:AiryFig3}}

\end{figure}

\begin{figure}[h]
\begin{centering}
\includegraphics[width=80mm]{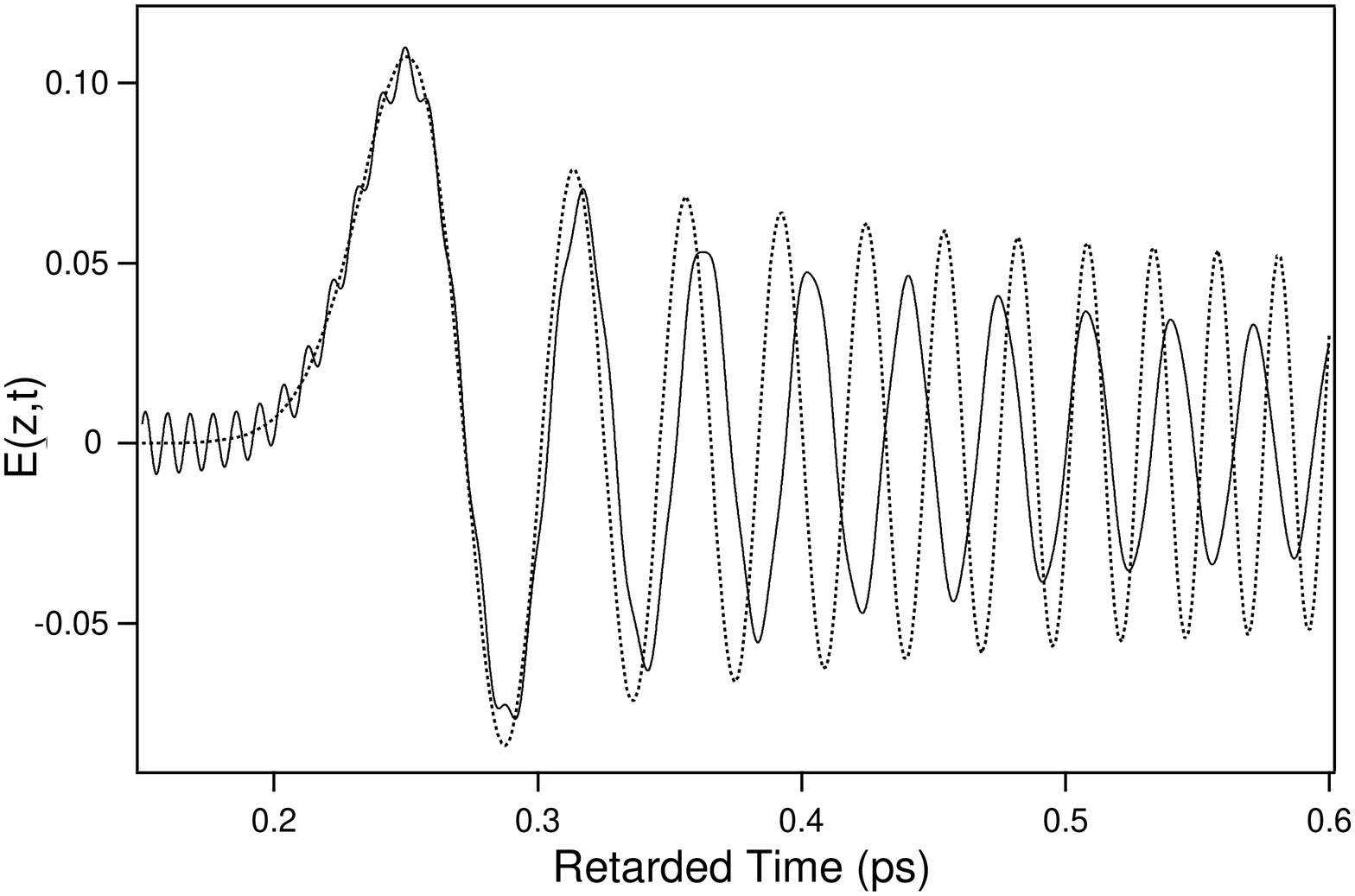} 
\par\end{centering}

\caption{Same as Fig.1 for the parameters of the figure 5 of \cite{ou10},
that is for $\omega_{0}=3.9\times10^{14}\mathrm{rad/s}$, $\omega_{c}=3.0\times10^{14}\mathrm{rad/s}$,
$\omega_{p}=3.05\times10^{12}\mathrm{rad/s}$ , $\delta=3.02\times10^{12}\mathrm{rad/s}$,
and $z=2.290\mathrm{m}$. The rapid oscillations of small amplitude
superposed upon the beginning of the Brillouin precursor are the end
of the Sommerfeld precursor.\label{fig:AiryFig5}}

\end{figure}

Although it abundantly refers to the theoretical results obtained
by the asymptotic method, the study reported in \cite{ou10} is mainly
numerical. All the simulations are made for $\omega_{0}=3.9\times10^{14}\mathrm{rad/s}$
\cite{re1} and $\omega_{c}=3.0\times10^{14}\mathrm{rad/s}$ in a
normal dispersion region. The singular and weak dispersion limits are
respectively attained when $\delta\ll\omega_{0}$ and $\omega_{p}^{2}\ll\delta\omega_{0}$
{[}see Eq. (\ref{eq:1}){]}. Oughstun \emph{et al.} emphasize that
these two limiting cases \textquotedblleft{}are fundamentally different
in their effects upon propagation\textquotedblright{} but, surprisingly
enough, they take for their simulations in the weak dispersion limit
a value of $\delta$ for which the singular limit nearly holds ($\delta<\omega_{0}/100$).
Consequently, \emph{mutatis mutandis}, the results obtained in the
two limits appears qualitatively similar. The steep rise of the Brillouin
precursor obtained in the singular limit (their Fig.3) and analyzed
as a distinguishing feature of this limit is only due to an unsuitable
time scale. As shown in our Fig.\ref{fig:AiryFig3}, obtained for
the same values of the parameters, the rise is quite progressive and
well reproduced by a Airy function.%
 This result, established by Brillouin himself in 1932 \cite{bri32},
is easily retrieved from Eqs (\ref{eq:1}), (\ref{eq:2}), and (\ref{eq:3}).
Anticipating that the beginning of the precursor involves frequencies
$\omega$ such that $\delta\ll\omega\ll\omega_{c}$, we use the approximate
relations 
\begin{equation}
n(\omega)\approx\left(1+\frac{\omega_{p}^{2}}{\omega_{0}^{2}}\right)^{1/2}+\frac{\omega^{2}\omega_{p}^{2}}{2\omega_{0}^{3}\left(\omega_{0}^{2}+\omega_{p}^{2}\right)^{1/2}}\label{eq:5}
\end{equation}
 and $\widetilde{E}(0,\omega)\approx1/\omega_{c}$. Introducing the new retarded time $t'=t-t_{b}$ with $t_{b}=\frac{z}{c}\left[\left(1+\omega_{p}^{2}/\omega_{0}^{2}\right)^{1/2}-1\right]$,
the transfer function then reads as $\exp\left[i\omega^{3}/\left(3b^{3}\right)\right]$
where
\begin{equation}
b=\omega_{0}\left[\frac{2c\left(\omega_{0}^{2}+\omega_{p}^{2}\right)^{1/2}}{3z\omega_{p}^{2}}\right]^{1/3}.\label{eq:6}
\end{equation}
We finally get 
\begin{equation}
E(z,t')=\frac{1}{2\pi\omega_{c}}\int_{-\infty}^{+\infty}\exp\left(\frac{i\omega^{3}}{3b^{3}}-i\omega t'\right)\mathrm{d}\omega=\frac{b}{\omega_{c}}\mathrm{Ai}(-bt')\label{eq:7}
\end{equation}
where $\mathrm{Ai}(x)$ is the Airy function \cite{re2}. As it appears
Fig.\ref{fig:AiryFig3}, this analytic expression perfectly fits not
only the rise of the precursor but also a significant number of its
subsequent oscillations. Equations (\ref{eq:6}) and (\ref{eq:7})
also provide a simple evidence of the $z^{-1/3}$ dependence of the
precursor amplitude on the propagation distance. Figure \ref{fig:AiryFig5}
shows the Brillouin precursor obtained in the conditions of the Figure
5 of \cite{ou10}, intended to illustrate the specific case of the
weak dispersion limit. As indicated previously, the conditions of
the singular limit are approximately met and this explains why the
first oscillation of the precursor and thus its peak amplitude are
well reproduced by Eq. (\ref{eq:7}). Note that the rise-time of the
precursor, proportional to $1/b$, is 6 times faster than in the previous
case. 

In their study of the weak dispersion limit, Oughstun \emph{et al}.
\cite{ou10} mention a \textquotedblleft{}curious difficulty in the
numerical FFT simulation of pulse propagation\textquotedblright{}
and, in order to overcome it, they develop at great length an equivalence
relation. In fact the difficulty is completely avoided in the retarded-time
picture and their equivalence relation then appears as an immediate
result. In the weak dispersion limit, the transfer function, as given
by Eq.(\ref{eq:2}), is indeed reduced to 
\begin{equation}
H(\omega)\approx\exp\left[-\frac{i\omega}{2c}\left(\frac{\omega_{p}^{2}z}{\omega^{2}-\omega_{0}^{2}+2i\delta\omega}\right)\right].\label{eq:8}
\end{equation}
 This only expression shows that, for given $\omega_{0}$ and $\delta$,
all the media having the same $\omega_{p}^{2}z$ (and thus the same
optical thickness at any reference frequency) are equivalent.

\begin{figure}[t]
\begin{centering}
\includegraphics[width=80mm]{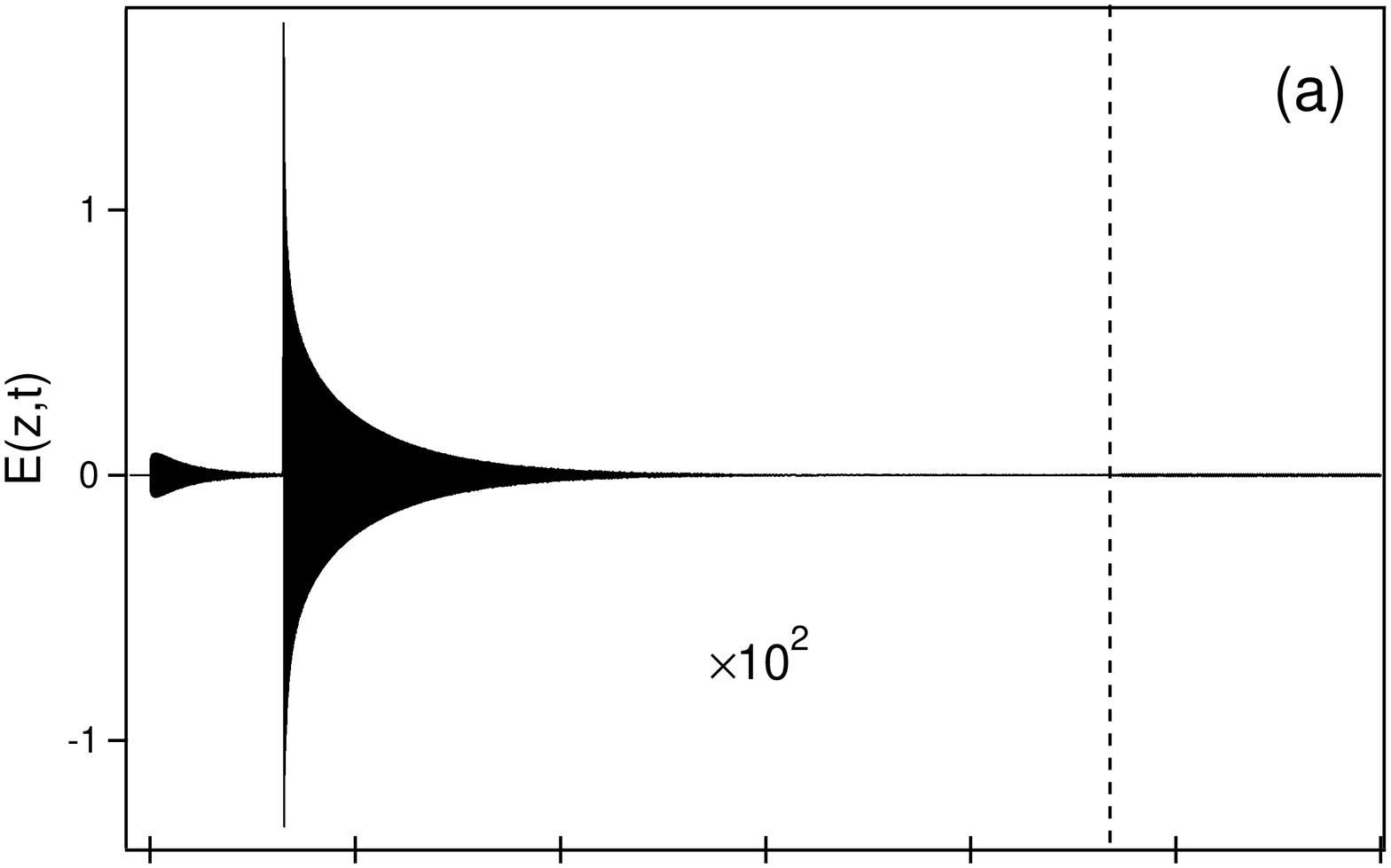} 
\par\end{centering}

\begin{centering}
\includegraphics[width=80mm]{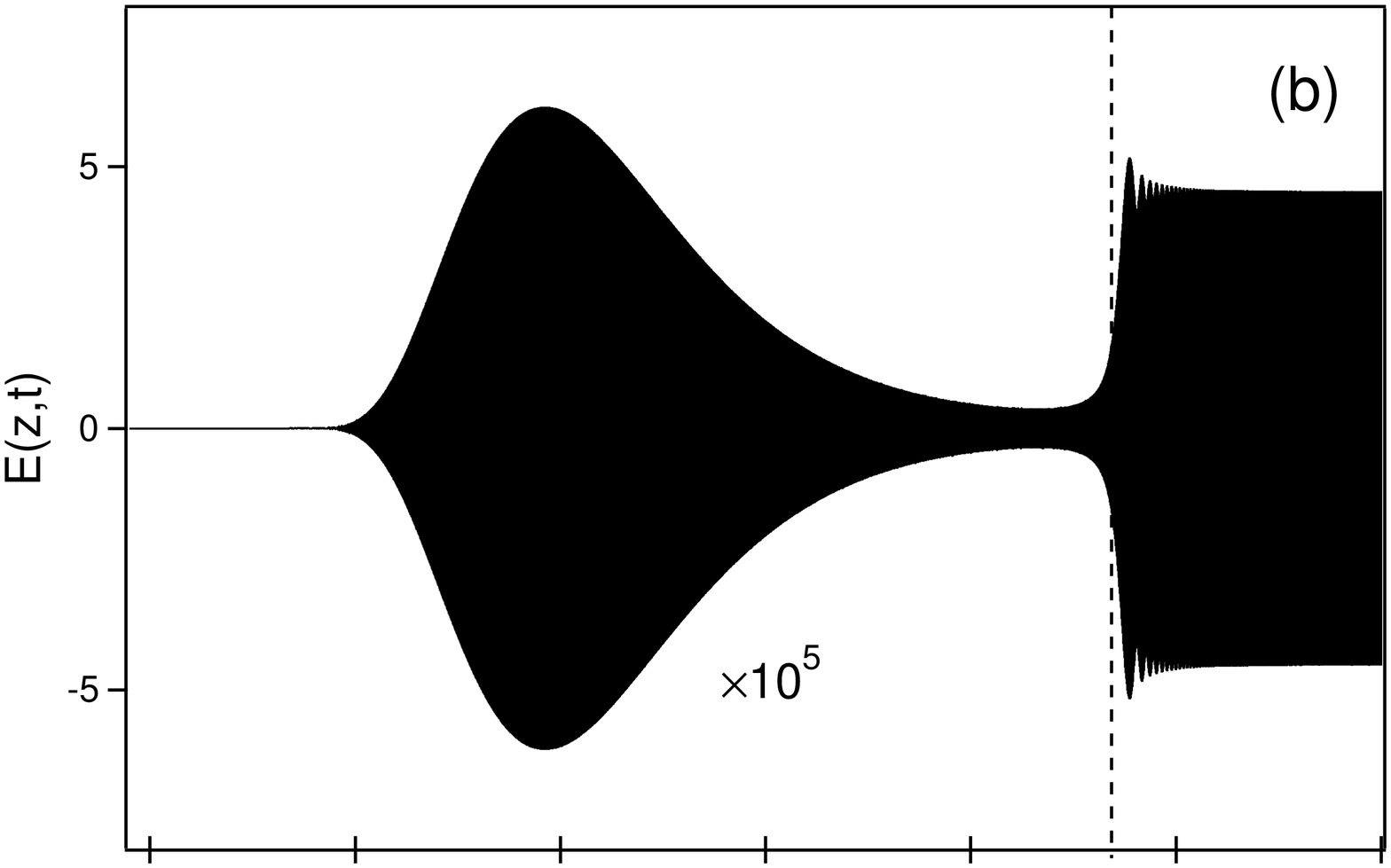} 
\par\end{centering}

\begin{centering}
\includegraphics[width=80mm]{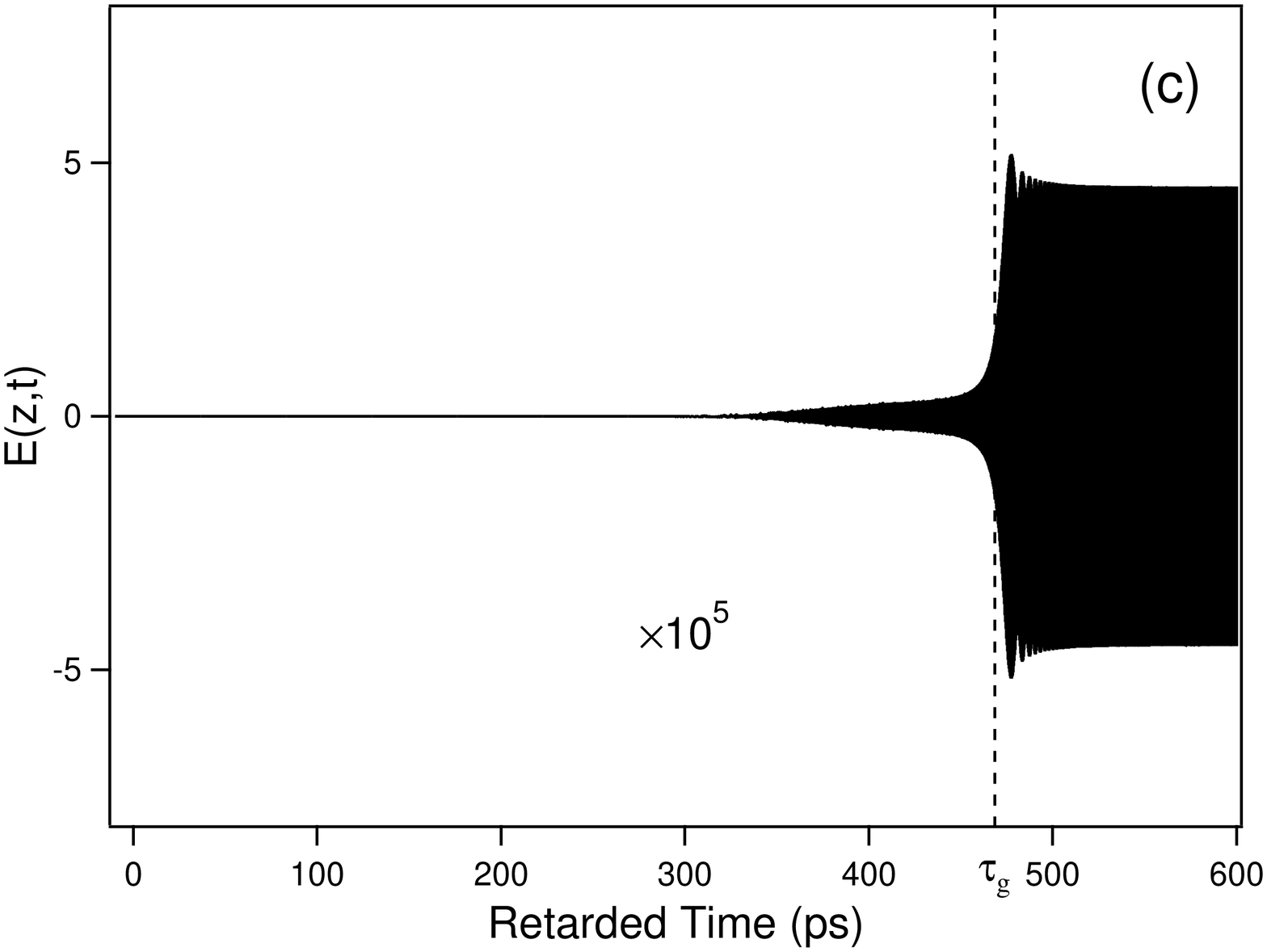} 
\par\end{centering}

\caption{Effect of the rise-time of the incident field on the transmitted field.
The parameters are those of the figure 3 of \cite{ou10}, that is $\omega_{0}=3.9\times10^{14}\mathrm{rad/s}$, $\omega_{c}=3.0\times10^{14}\mathrm{rad/s}$,
$\omega_{p}=3.05\times10^{14}\mathrm{rad/s}$ , $\delta=3.02\times10^{10}\mathrm{rad/s}$,
and $z=7.232\times10^{-2}\mathrm{m}$. The vertical dashed line indicates
the group delay $\tau_{g}=\frac{\mathrm{d}\varphi}{\mathrm{d}\omega}\left|_{\omega=\omega_{c}}\right.$
where $\varphi$ is the argument of $H(\omega)$ ($\tau_{g}=471\mathrm{ps}$).
$\tau_{g}$ fixes the arrival of the \textquotedblleft{}main field\textquotedblright{}
whose amplitude, equal to $\exp\left(-10\right)\approx4.5\times10^{-5}$,
obviously does not depend on the rise-time. The different curves are
obtained for (a) $T_{r}=0$, (b) $T_{r}=\tau/500$, and (c) $T_{r}=\tau/100$
where $\tau=1/\delta$ is the relaxation or damping time of the medium.\label{fig:Finit-Rise-Time}}

\end{figure}

In a real experiment, the incident field is obviously turned on in
a finite time. Oughstun \emph{et al}. state that their \textquotedblleft{}results
will remain valid for a non instantaneous turn-on signal provided
that the signal turn-on time $T_{r}$ is faster than the characteristic
relaxation time $1/\delta$\textquotedblright{}. This is grossly false.
Figure \ref{fig:Finit-Rise-Time} shows the results of a simulation
performed for the parameters of their figure 3 and $10-90\%$ turn-on
times (a) $T_{r}=0$, (b) $T_{r}=\tau/500$, and (c) $T_{r}=\tau/100$
where $\tau=1/\delta$ is the relaxation or damping time ($\tau\approx33\mathrm{ps}$).
The rise of the incident field is modelled by replacing the step-function
$\Theta(t)$ appearing in the idealized form of Eq. (\ref{eq:4})
by $f(t)=\frac{1}{2}\left[1+\mathrm{erf}(\xi t)\right]$ where $\mathrm{erf}(x)$
designates the error function. The corresponding rise-time is $T_{r}\approx1.8/\xi$
and $f(t)\rightarrow\Theta(t)$ when $\xi\rightarrow\infty$. We see
that the effect of the rise-time is actually catastrophic. For a rise-time
as fast as $\tau/500$ ($\approx66\mathrm{fs}$ ), the Sommerfeld
precursor is absent. The maximum of the Brillouin precursor is significantly
time-delayed and its amplitude $A_{B}$ is reduced by a factor of
about 300, becoming comparable to the amplitude $\mathrm{e}^{-10}$
of the \textquotedblleft{}main field\textquotedblright{}. The reduction
factor obviously depends on the optical thickness and, consequently,
the power law $A_{B}\propto z^{-1/3}$, obtained for $T_{r}=0$, breaks
down. Finally the curve (c) of Fig.\ref{fig:Finit-Rise-Time} shows
that both precursors practically vanish for $T_{r}=\tau/100$ . By means
of other simulations, we find that the Sommerfeld precursor is very
attenuated as soon as $T_{r}>\tau/5000\approx2/\omega_{c}$ and that
a correct reproduction of both precursors as obtained for $T_{r}=0$
requires that $T_{r}\leq1/\omega_{c}$.  This condition results from the fact that the Sommerfeld (Brillouin) precursor mainly involves frequencies $\omega\gg\omega_{c}$  ($\omega\ll\omega_{c}$) and thus is excited by the corresponding frequencies contained in the spectrum of the incident field. From the expression of $f(t)$, it
is easily shown that the effect of a finite rise-time $T_{r}$ ($T_{r}\approx1.8/\xi$
) is to divide $\widetilde{E}(0,\omega)$ by $\exp\left[\omega^{2}/\left(4\xi^{2}\right)\right]$
when $\omega/\omega_{c}\rightarrow\infty$ and by $\exp\left[\omega_{c}^{2}/\left(4\xi^{2}\right)\right]$
when $\omega/\omega_{c}\rightarrow0$. These expressions explain why the Sommerfeld precursor is much more affected by the rise-time effects than the Brillouin precursor and enable one to predict that, for $T_{r}=1/\omega_{c}$ ($\xi\approx1.8 \omega_{c}$), the amplitude  $A_{B}$ of the Brillouin precursor will be about  $7.5\%$ below that obtained for $T_{r}=0$ (result confirmed by an exact numerical calculation).

The observation of a signal close to that shown on the figure 3 of
\cite{ou10} requires not only that the rise-time of the incident
field does not exceed $1/\omega_{c}$ ($3.3\mathrm{fs}$) but also
that its subsequent amplitude remains nearly constant during a time
that would be more than four orders of magnitude longer. The fulfilment
of this double condition, either with a pulsed laser or with a continuous
wave laser followed by a modulator, appears to be quite unrealistic
from an experimental viewpoint. More generally, due to such temporal
constraints, we don\textquoteright{}t see how the Brillouin precursor
could be actually used in optics as a tool for imaging through a dense
medium, opaque at the carrier frequency. Anyway, the
problem of optical precursors is of fundamental interest from a theoretical
viewpoint, even if the experiment imagined by Sommerfeld and Brillouin
may appear as a \emph{gedankenexperiment}. In this spirit, we are
examining what the precursors become when the incident field $E(0,t)=\Theta(t)\sin(\omega_{c}t)$,
constantly considered in the theoretical papers, is replaced by $E(0,t)=\Theta(t)\cos(\omega_{c}t)$.
The true discontinuity of the incident field at the initial time then
leads to radically new effects that we will discuss in a forthcoming
paper.


\begin{thebibliography}{4}
\bibitem{re1} Contrary to what is stated in \cite{ou10}, this (circular)
frequency $\omega_{0}$ is not equal to the frequency of the resonance
involved in the experiment on cold potassium atoms but is $2\pi$
times smaller.
\bibitem{re2} Note that the Airy function $\mathcal{A}(x)$ used
by Brillouin in \cite{bri32} differs from the standard one $\mathrm{Ai}(x)$
used in the present comment, with $2\pi\mathrm{Ai}(-x)=3^{1/3}\mathcal{A}(3^{1/3}x)$
. Peak amplitudes of $\mathcal{A}(x)$ and $\mathrm{Ai}(-x)$ are
respectively 2.33 and 0.536.
\end{thebibliography}
\end{document}